\def \nfpkois {685~}
\def \nkois {5,785~}
\def \nkebs {2,604~}
\def \nkebstot {4,403~}
\def \ngebstot {2,044~}
\def \nuniqsource {409~}
\def \nbastards {31~}
\def \nmultikois {1,661~}
\def \nfpmultis {40~}

\documentclass[iop,apj]{emulateapj}
\begin{document}

\title{Contamination in the \emph{Kepler} Field. Identification of \nfpkois KOIs as False Positives Via Ephemeris Matching Based On Q1-Q12 Data}
\author{Jeffrey L. Coughlin\altaffilmark{1,2,3}, Susan E. Thompson\altaffilmark{2,3}, Stephen T. Bryson\altaffilmark{3}, Christopher J. Burke\altaffilmark{2,3}, Douglas A. Caldwell\altaffilmark{2,3}, Jessie L. Christiansen\altaffilmark{5}, Michael R. Haas\altaffilmark{3}, Steve B. Howell\altaffilmark{3}, Jon M. Jenkins\altaffilmark{2,3}, Jeffery J. Kolodziejczak\altaffilmark{4}, Fergal R. Mullally\altaffilmark{2,3}, and Jason F. Rowe\altaffilmark{2,3}}

%\journalinfo{} and \submitted{}
\journalinfo{Accepted for Publication in the Astronomical Journal - Feb. 21, 2014}
 \submitted{}

\altaffiltext{1}{jeffrey.l.coughlin@nasa.gov}
\altaffiltext{2}{SETI Institute, 189 Bernardo Ave, Mountain View, CA 94043, USA}
\altaffiltext{3}{NASA Ames Research Center, M/S 244-30, Moffett Field, CA 94035, USA}
\altaffiltext{4}{NASA Marshall Space Flight Center, Huntsville, AL 35812, USA}
\altaffiltext{5}{NASA Exoplanet Science Institute, California Institute of Technology, Pasadena CA 91125, USA}

\begin{abstract}
The \emph{Kepler} mission has to date found almost 6,000 planetary transit-like signals, utilizing three years of data for over 170,000 stars at extremely high photometric precision. Due to its design, contamination from eclipsing binaries, variable stars, and other transiting planets results in a significant number of these signals being false positives. This directly affects the determination of the occurrence rate of Earth-like planets in our Galaxy, as well as other planet population statistics. In order to detect as many of these false positives as possible, we perform ephemeris matching among all transiting planet, eclipsing binary, and variable star sources. We find that \nfpkois \emph{Kepler} Objects of Interest --- 12\% of all those analyzed --- are false positives as a result of contamination, due to \nuniqsource unique parent sources. Of these, 118 have not previously been identified by other methods. We estimate that $\sim$35\% of KOIs are false positives due to contamination, when performing a first-order correction for observational bias. Comparing single-planet candidate KOIs to multi-planet candidate KOIs, we find an observed false positive fraction due to contamination of 16\% and 2.4\% respectively, bolstering the existing evidence that multi-planet KOIs are significantly less likely to be false positives. We also analyze the parameter distributions of the ephemeris matches and derive a simple model for the most common type of contamination in the \emph{Kepler} field. We find that the ephemeris matching technique is able to identify low signal-to-noise false positives that are difficult to identify with other vetting techniques. We expect false positive KOIs to become more frequent when analyzing more quarters of \emph{Kepler} data, and note that many of them will not be able to be identified based on \emph{Kepler} data alone.
\end{abstract}


\section{Introduction}
\label{introsec}

NASA's \emph{Kepler} mission is a 0.95 meter aperture, optical (420 - 915nm), space-based telescope that was launched in 2009 with the primary goal of determining the occurrence rate of Earth-sized planets in our Galaxy \citep{Borucki2010a}. It is able to achieve this goal by employing 42 CCDs to constantly observe $\sim$170,000 stars over a field of view (FoV) of 115 square degrees \citep{Koch2010}, searching for the periodic drops in brightness that occur when planets transit in front of their host stars. A photometric precision of $\sim$40~ppm is attained on a 6-hour timescale for a 12$^{\rm th}$ magnitude star \citep{Christiansen2012}. With several years of data, it is thus capable of detecting signals (depending on the period) as low as several ppm.

This extreme photometric sensitivity for so many stars comes at a price. Given the very large field of view, each pixel spans 3.98$\arcsec$ \citep{Koch2010}, and while the telescope is at an optimal focus that minimizes the point spread function (PSF) across the entire focal plane, local regions have non-optimal PSFs. While spacecraft pointing is precise to $\sim$0.2$\arcsec$ (0.05 pixels) over a quarter ($\sim$90 day period), the amount of stellar aberration introduced by the spacecraft's velocity varies across the field of view. This results in differential velocity aberration and thus the shifting of stellar positions on the detector by as much as 2.4$\arcsec$ (0.6 pixels) over a quarter \citep{Jenkins2010b}. The \emph{Kepler} pixel response function (PRF) is the combination of the telescope's point spread function (PSF), the CCD pixel resolution, and the spacecraft's pointing jitter over each quarter. The combination of these effects result in a PRF with a 95\% encircled flux radius of $\sim$16--28$\arcsec$ (4--7 pixels), with an increasingly asymmetric PRF towards the edge of the field of view \citep{Bryson2010a}.

Since the \emph{Kepler} spacecraft is of Schmidt design, it features a fused-quartz Schmidt corrector plate that compensates for the spherical aberration induced by the primary mirror. Each CCD also has an individual field flattener lens to map the spherical telescope image onto the flat CCDs. These multiple reflecting surfaces result in a significant amount of stray light in the system, compared to \emph{Kepler's} extreme photometric precision. Additionally, the large number of CCDs, which are all read out simultaneously, allows for a significant chance of electronic interaction.

The large PRF and multiple optical and electronic components allow for significant contamination to occur. Contamination is defined as light in the photometric aperture of a target that does not actually originate from that target. If the extra light comes from a variable source then that variable signal will be observed in the target, with a reduced amplitude due to dilution. For example, a target that has no intrinsic variability could be contaminated by an EB with an intrinsic 50\% eclipse depth, which manifests itself as a 0.5\% transit-like signal in the target. This transit signal would thus be a false positive (FP), because the signal does not actually originate from the target.

Throughout this paper we will refer to the contaminating source as the ``parent" and the contaminated target as the ``child", e.g., the EB is the parent and the transit-like signal is the child in the previous example. Multiple children caused by the same parent will be referred to as ``siblings". Due to on-board storage and bandwidth constraints, only 5.44 million of the 96 million pixels (5.4\%) \citep{Bryson2010b}, or 170,000 of the 500,000 stars (34\%) with \emph{Kepler} magnitude brighter than 16.0 \citep{Batalha2010}, are downloaded from the spacecraft. As a result, a child and its siblings may be observed while the parent is not. In these cases, we refer to the children as ``bastards" (since the term ``orphan" is already used elsewhere in the \emph{Kepler} literature.)

The traditional method of identifying FPs due to contamination has been to examine the pixel-level data and identify exactly which pixels in and around the target contain the transit signal. A thorough review of this technique is given by \citet{Bryson2013}, and its application to eliminating FPs from planet candidate catalogs is shown in \citet{Borucki2011a}, \citet{Borucki2011b}, \citet{Batalha2013}, \citet{Burke2013}, and \citet[][in preparation]{Rowe2014}. Essentially, if the target is a FP due to contamination, the pixel location of the transit signal will not coincide with the target's flux distribution. Synergistically, if the parent is an EB with a deep secondary eclipse, it is also sometimes possible to see this eclipse in the light curve of the child, and thus confirm the child as a FP. These techniques work well for cases where the child is close to the parent, and the transit signal has a high signal-to-noise ratio (SNR). However, if the parent is far away from the child, the contaminating flux may be too diffuse to definitively determine that the transit does not occur on the target. If the SNR is too low there might not be enough signal in each individual pixel to be able to tell which pixels contain the transit signal nor see a secondary eclipse.

An alternative to examining the pixel-level data is to see if two targets have matching ephemerides, i.e., they have the same period and epoch, which indicates at least one of them is a FP due to contamination. There have been some previous, limited attempts to discover or confirm FPs due to contamination via ephemeris matching. \citet{Batalha2013} found $\sim$25 planetary candidates (PCs) to actually be FPs by examining candidates that were within 20$\arcsec$ of another PC or EB) and had matching periods. The Planet Hunters citizen science project employs many volunteers to manually inspect \emph{Kepler} light curves for transiting planets, and has been successful in finding many candidates missed by other methods \citep[e.g.,][]{Fischer2012,Schwamb2012}. Discussion and analysis of planet candidates takes place on the discussion forums\footnote{\url{http://talk.planethunters.org}}, and many volunteers have identified candidates to be period matches to nearby eclipsing binaries or transiting planets. Furthermore, individuals have contributed to the \emph{Kepler} project with individual period matching lists \citep[e.g.,][private communication]{Shporer2013}.

The goal of this paper is to perform a comprehensive and well-documented search for contamination via ephemeris matching, utilizing the latest planet and eclipsing binary catalogs, with physically plausible constraints. We focus on the \emph{Kepler} Objects of Interest (KOIs), which are a catalog of transit-like signals in the \emph{Kepler} data. In \S\ref{contammethsec} we describe the different mechanisms by which stars can contaminate each other with the \emph{Kepler} instrument. In \S\ref{searchsec} we describe how we compiled our input catalogs, searched for ephemeris matches, and compiled our final table of FP KOIs.  Finally in \S\ref{discusssec} we discuss the prevalence of each method of contamination, analyze the parameter distributions of FP KOIs and their parents, compare our results to previous work identifying false positives, and anticipate future work in the field.

\section{Methods of Contamination}
\label{contammethsec}

In this section, we distinguish between the different physical mechanisms of contamination. We discuss the four currently known mechanisms and describe each. Every FP KOI that will be presented in \S\ref{searchsec} is assigned to one of these four categories.

\subsection{Direct PRF}
\label{directprfsec}

Direct PRF contamination occurs when the PRF of two stars overlap, such that light from a parent star is directly included in one or more pixels that comprise the optimal photometric aperture of a child star. Due to the aforementioned spacecraft design specifications listed in \S\ref{introsec}, this is quite common across the entire field of view. The range at which the PRF wings are above the CCD noise limit can extend to over a hundred arcseconds for bright stars (see \S\ref{fpkoiprops}).

Another effect that we choose to include in this category is reflection off the spacecraft's field flattener lenses. Light can reflect off a CCD, then off the lens above it, and back onto the CCD, resulting in a very large out of focus ghost image \citep{VanCleve2009}. This effect is usually only seen for very bright stars, as the multiple reflections involved quickly reduce the light level to below the CCD read noise level. The resulting relative per-pixel signal strength was estimated to be 10$^{-5}$ prelaunch \citep{Caldwell2010}. However, when it is measurable, this effect greatly extends the wings of a bright star's PRF and allows it to contaminate to over one thousand arcseconds, as the resulting image is spread over thousands of pixels \citep{Caldwell2010}.

\subsection{Antipodal Reflection}
\label{reflectionsec}

The \emph{Kepler} spacecraft is of Schmidt design, and thus features a fused-quartz Schmidt corrector plate that compensates for the spherical aberration induced by the primary mirror. Light is able to reflect off a CCD, then off the Schmidt corrector plate, and back onto another CCD. Due to the optical design, the location of the resulting ghost image is antipodal to the parent source, i.e., on the opposite side of the FoV with respect to its center. While the resulting ghost image is not identical to the parent's PRF, it does have a similar value for full-width at half-max. The signal strength of the ghost image was estimated prelaunch to be 10$^{-3.4}$ relative to the parent \citep{Caldwell2010}.

\subsection{CCD Crosstalk}
\label{crosstalksec}

\emph{Kepler's} 42 CCDs are paired into 21 modules, each of which has 4 outputs. Each output reads out half of a CCD, and thus there are a total of 84 channels, one for each output. Electronic crosstalk is a physical effect where the electronic signal in one wire can electromagnetically induce that signal in other bundled, parallel wires. With respect to \emph{Kepler}, each of the four CCD outputs on each module are bundled and read out simultaneously. Thus, a signal from pixel (x,y) on output 1 can induce an identical signal on pixel (x,y) on outputs 2, 3, and 4, though at a lower amplitude. The severity of the resulting crosstalk can vary greatly depending on which outputs are interacting, with crosstalk coefficients ranging from -10$^{-2}$ to 10$^{-2}$ \citep{Caldwell2010}. 

The cumulative effect of crosstalk is that a given parent star will create a ghosted image of itself at the same pixel position on 3 other outputs on the same module. Any time-varying signal that is present in the star, e.g., an eclipsing binary, will also be induced on the other 3 outputs. If a star on one of those other outputs happens to have an aperture that includes pixels containing the crosstalk signal, that star will be a contaminated child. The result is apparent variability in the target child star that matches the variability from the contaminating parent. 

As the spacecraft rotates every quarter ($\sim$90 days), a given star will fall on one of four different CCDs, one for each of the four seasons. As each CCD has its own particular physical characteristics and wiring variations, the amount of crosstalk can vary greatly in different seasons. Also, because targets do not fall on the same pixels each quarter, due to imperfect CCD registration, a child may not share the same (x,y) coordinates as its parent in every quarter. Both of these effects result in drastically varying levels of crosstalk contamination each quarter.

To understand the importance of this crosstalk effect on the KOI population, we can evaluate each KOI for the potential of being contaminated by crosstalk. Only targets that share the same pixels as a bright and highly variable target, on an adjacent outputs, can have variability caused by crosstalk. Using the Full Frame Images (FFIs), the measured crosstalk coefficients, and the optimal apertures, we can evaluate the size of the crosstalk flux for any target. Since \emph{Kepler} does not obtain a time series for all pixels, we do not necessarily know if the parent is variable.  However, if we assume that the contaminating flux varies by less than 100~percent, we can determine the largest transit on a target that could be caused by crosstalk.  Given the transit depth of the Q1-Q12 KOIs, we calculate that less than 9\% of KOIs have at least one quarter of data whose transit could be caused by crosstalk. In fact, the true number of KOIs caused by crosstalk must be much less than this since very few sources have variability as large as 100~percent. Thus, we expect to find very few KOIs impacted by this method of contamination, as is evident in the results of the ephemeris matching presented in this paper (see \S\ref{searchsec}).

\subsection{Column Anomaly}
\label{columncontamsec}

In the course of our search, we noticed a new type of contamination that had not been previously anticipated. Apparently the signal from a parent can contaminate a child that lies on approximately the same column of that CCD, up to the entire range of the CCD. One would initially suspect saturation, as the excess charge from a saturated star overflows into neighboring rows along the same column. While a few cases might be due to saturation, most of the parents and children are in fact not saturated, and both have been observed to be as faint as $\sim$15$^{\rm th}$ magnitude. While charge transfer efficiency, smear correction, signal quantization, and many other mechanisms are being investigated, a single physical mechanism to explain the column anomaly has not yet been found.

\section{The Search for Ephemeris Matches}
\label{searchsec}

\subsection{Catalog Compilation}
\label{catcompsubsec}

We employed the following sources to create catalogs of transiting planets, eclipsing binaries, and other variable stars in the \emph{Kepler} field of view:

\begin{itemize}

\item The list of \nkois \emph{Kepler} objects of interest (KOIs), ranging from KOI 1.01 to 4914.01, available at the NASA Exoplanet Archive\footnote{\url{http://exoplanetarchive.ipac.caltech.edu}} as of December 18, 2013. These include KOIs detected utilizing up to 12 quarters of data \citep[][in preparation]{Rowe2014}, as well as previous catalogs \citep{Burke2013,Batalha2013,Borucki2011b,Borucki2011a}. Although the previous catalogs contained only transiting planet-like signals, this most recent catalog contains both transiting planets and eclipsing binaries.

\item The \emph{Kepler} Eclipsing Binary Catalog list of \nkebs ``true" EBs found via \emph{Kepler} data as of December 18, 2013.\footnote{\url{http://keplerebs.villanova.edu}} The compilation of the catalog and derivation of the fit parameters are described in \citet[][in preparation]{Kirk2013}. Previous versions of this catalog are described in \citet{Slawson2011} and \citet{Prsa2010}.

\item J.M. Kreiner's up-to-date database of ephemerids of ground-based eclipsing binaries as of December 18, 2013.\footnote{\url{www.as.up.krakow.pl/ephem}} Data compilation and parameter derivation are described in \citet{Kreiner2004}.

\item Ground-based eclipsing binaries found via the TrES Survey as detailed in \citet{Devor2008a}.

\item The General Catalog of Variable Stars (GCVS) list of all known ground-based variable stars, published December 2013.\footnote{\url{www.sai.msu.su/gcvs/gcvs}} This catalog includes both eclipsing binaries and other periodic variable stars, such as pulsators. Catalog compilation is described by \citet{Samus2013}.

\end{itemize}

From these sources, we created three separate catalogs to perform the ephemeris matching: a KOI catalog, a \emph{Kepler}-based EB (KEB) catalog, and a ground-based EB (GEB) catalog. The GEB catalog was trimmed in RA/Dec space to include only those stars that fell within 20 degrees of the \emph{Kepler} FoV center (19h~22m~40s, +44$\degr$~30\arcmin~00$\arcsec$), which ensures that all on-sky CCDs are covered by a few extra degrees.  For each eclipsing binary we designated the primary eclipse by appending ``-pri" to the name, and if the time of minimum of the secondary eclipse was given, we created a separate entry for the secondary eclipse and appended a ``-sec". In the case of the GCVS catalog, if a secondary eclipse exists the depth is given, but not the time of minimum, and thus we assumed circular orbits for GCVS secondary times of minimum. 

We employed the \emph{Kepler} Input Catalog (KIC) \citep{Brown2011} and the \emph{Kepler} Characteristics Table\footnote{\url{http://archive.stsci.edu/pub/kepler/catalogs/}} to obtain additional parameters for each object. While all KOIs and KEBs already have KIC numbers, we had to assign KIC numbers to each GEB via coordinate matching. We utilized the KIC search page\footnote{\url{http://archive.stsci.edu/kepler/kic10/search.php}} to find the closest KIC star to each GEB within 0.02 arcminutes. Only about half of the GEBs had matching KIC stars, which is not surprising as a 20 degree radius more than covers the \emph{Kepler} FoV by a few degrees.

For each object, if the data existed, we gathered the values of right ascension, RA, declination, Dec, period, $P$, time of minimum, $T$, depth of transit/eclipse, $D$, \emph{Kepler} magnitude, $m_{\rm kep}$, and the CCD channel number, $chan$, module number, $mod$, output number, $out$, row number, $row$, and column number, $col$, for each season. In the cases where GEBs occurred in both the GCVS catalog as well as the \citet{Kreiner2004} catalog, we chose to use period and epoch values from \citet{Kreiner2004} as they are generally more up-to-date and accurate. For the GEBs without KIC IDs, we made the assumption that $m_{\rm kep}$ $\approx$ $m_{\rm V}$. GEBs that do not fall on a CCD do not have the associated CCD location information.

In total there were \nkois entries for the KOI catalog, \nkebstot entries for the KEB catalog, and \ngebstot entries for the GEB catalog. We list each of these 12,232 entries in Table~\ref{paramtab} along with their KIC number, Period, Epoch, Depth, \emph{Kepler} Magnitude, RA, Dec, and Season 0 Channel, Module, Output, Row, and Column numbers.

\begin{deluxetable*}{rcrcrrccrrrr}
\tablecolumns{12}
\tablewidth{0in}
\tablecaption{KOI, KEB, and GEB properties}
\tablehead{\colhead{Name} & \colhead{KIC ID} & \colhead{P} & \colhead{T} & \colhead{D} & \colhead{$m_{\rm kep}$} & \colhead{RA} & \colhead{Dec.} & \colhead{Mod$_{0}$} & \colhead{Out$_{0}$} & \colhead{Row$_{0}$} & \colhead{Col$_{0}$}\\ & & \colhead{(Days)} & \colhead{(BJD-2.4E6)} & \colhead{(ppm)} & & \colhead{(Hours)} & \colhead{(Deg)} & & & & }
\startdata
\\
\cutinhead{Kepler Objects of Interest (KOIs)}\\
1.01 & 011446443 &    2.47061317 & 54955.762566 &  14284 & 11.338 & 19.120565 & 49.316399 & 23 & 1 & 848 & 618\\
2.01 & 010666592 &    2.20473537 & 54954.357802 &   6713 & 10.463 & 19.483152 & 47.969521 & 19 & 2 & 614 & 1047\\
3.01 & 010748390 &    4.88780026 & 54957.812537 &   4323 &  9.174 & 19.847290 & 48.080853 & 15 & 3 & 819 & 556\\
\nodata & \nodata & \nodata & \nodata & \nodata & \nodata & \nodata & \nodata & \nodata & \nodata & \nodata & \nodata\\
\cutinhead{Kepler Eclipsing Binaries (KEBs)}\\
001026032-pri & 001026032 &    8.46043800 & 54966.773813 &  71700 & 14.813 & 19.402939 & 36.729271 & 2 & 4 & 201 & 97\\
001026032-sec & 001026032 &    8.46043800 & 54971.229080 &  27100 & 14.813 & 19.402939 & 36.729271 & 2 & 4 & 201 & 97\\
001026957-pri & 001026957 &   21.76130600 & 54956.017106 &   1200 & 12.559 & 19.416965 & 36.743610 & 2 & 4 & 130 & 232\\
\nodata & \nodata & \nodata & \nodata & \nodata & \nodata & \nodata & \nodata & \nodata & \nodata & \nodata & \nodata\\
\cutinhead{Ground-based Eclipsing Binaries (GEBs) With KICs}\\
AG-Cyg-pri & 001476573 &  296.30000000 & 34240.000000 & 995214 & 13.518 & 19.892700 & 37.043208 & \nodata & \nodata & \nodata & \nodata\\
AH-Cyg-pri & 005048397 &  112.00000000 & 29445.000000 & 748811 &  9.610 & 20.010200 & 40.180065 & \nodata & \nodata & \nodata & \nodata\\
AW-Dra-pri & 011802860 &    0.68719410 & 36075.200000 & 841511 & 13.053 & 19.013332 & 50.092018 & 23 & 4 & 192 & 959\\
\nodata & \nodata & \nodata & \nodata & \nodata & \nodata & \nodata & \nodata & \nodata & \nodata & \nodata & \nodata\\
\cutinhead{Ground-based Eclipsing Binaries (GEBs) Without KICs}\\
AA-Lyr-pri & \nodata &    1.07333900 & 52500.200000 & 424560 & 13.400 & 19.096917 & 29.078889 & \nodata & \nodata & \nodata & \nodata\\
AA-Lyr-sec & \nodata &    1.07332900 & 29734.100000 &  87989 & 13.400 & 19.096917 & 29.078889 & \nodata & \nodata & \nodata & \nodata\\
AA-Vul-pri & \nodata &  439.10000000 & 33913.000000 & 980945 & 13.700 & 19.840583 & 28.188889 & \nodata & \nodata & \nodata & \nodata\\
\nodata & \nodata & \nodata & \nodata & \nodata & \nodata & \nodata & \nodata & \nodata & \nodata & \nodata & \nodata
\enddata
\tablecomments{Table~\ref{paramtab} is published in its entirety in the electronic edition of the Astronomical Journal. A portion is shown here for guidance regarding its form and content.}
\label{paramtab}
\end{deluxetable*}

\subsection{Matching Criterion}
\label{matcritsubsec}

There are $\sim$54 million unique combinations when comparing the KOIs to themselves, to the KEB catalog, and to the GEB catalog. The quality of the period and epochs can significantly vary, and it is possible for the listed period to be an integer multiple of the ``true" period. We thus require precise yet flexible matching criteria to identify statistically significant ephemeris matches among the myriad of possible combinations.

For matching object A to object B, where object A has period $P_{A}$ and epoch $T_{A}$, and object B has $P_{B}$ and $T_{B}$, the following parameters were computed:

\begin{equation}
\Delta P = \frac{P_{A}-P_{B}}{P_{A}}
\qquad
\Delta T = \frac{T_{A}-T_{B}}{P_{A}}
\end{equation}

\begin{eqnarray}
\Delta P^{\prime} = abs(\Delta P - int(\Delta P))\\
\nonumber \Delta T^{\prime} = abs(\Delta T - int(\Delta T)) 
\end{eqnarray}

\noindent where $int()$ rounds a number to the nearest integer (e.g., 3.99 $\rightarrow$ 4, 4.01 $\rightarrow$ 4, -3.99 $\rightarrow$ -4 and -4.01 $\rightarrow$ -4) and $abs()$ yields the absolute value. 

These equations describe the fractional difference in period and epoch with respect to the period of object A. For perfect matches where the period and epoch are either identical or perfect ratios, $\Delta P^{\prime}$ = $\Delta T^{\prime}$ = 0.0. In order to easily comprehend a large range of small fractional values, we convert these fractional values into sigma values via:

\begin{eqnarray}
\sigma_{P} = \sqrt{2}\cdot\textrm{erfcinv}(\Delta P^{\prime})\\
\nonumber \sigma_{T} = \sqrt{2}\cdot\textrm{erfcinv}(\Delta T^{\prime})
\end{eqnarray}

\noindent where erfcinv() is the inverse complementary error function. For example, a $\Delta P^{\prime}$ value of 0.0027 means the periods of two objects, after accounting for any possible period ratios, only differ by 0.27\%, or alternatively stated agree to 99.73\%, which corresponds to a 3.0$\sigma$ match. Smaller values of $\Delta P^{\prime}$ will result in larger sigma values, with $\sigma\rightarrow\infty$ as $\Delta P^{\prime}\rightarrow$ 0.

We note that these equations allow for any integer period ratio, e.g., 1:1, 2:1, 42:1, etc, when $P_{B} > P_{A}$. They also allow for any offset in the time of minimum by integer values of the period, e.g., two objects each with periods of 2.0 days, but listed time of minima of 136.3 and 138.3 days. In general, shorter period objects have more precisely determined periods and epochs than longer period objects, as there are more transit/eclipse events in a given time range. Since these equations are based on the fractional differences in period and epoch between two objects, they also naturally require shorter period objects to match to a higher absolute timing than longer period objects.

In order to ensure only statistically significant, physically plausible matches were found, all three of the following criteria had to be met to establish a match:

\begin{enumerate}

\item The two objects could not have the same KIC ID.

\item The two objects had to satisfy at least one of the following conditions: 

    \begin{enumerate}
    
    \item Have a separation distance of less than $d_{\rm max}$ arcseconds of each other, where
    \begin{equation}
    \label{disteq}
    d_{\rm max}(\arcsec) = 50\cdot\sqrt{10^{6}\cdot 10^{-0.4 \cdot m_{\rm kep}}+1}
    \end{equation}

\noindent and where the magnitude of the brighter source is used for $m_{\rm kep}$.  

    \item Be located on equidistant, opposite sides of the FoV center within a 50$\arcsec$ (12.5 pixel) tolerance.
    
    \item Be located on the same CCD module and be within 10 pixels of either the same row or column value, for any of the 4 quarters.

   \end{enumerate}

\item Both objects had to match to better than minimum $\sigma$ values of $\sigma_{P,\rm min}$ and $\sigma_{T,\rm min}$ (defined below) either from matching object A to B, or B to A.

\end{enumerate}

\begin{figure}
\centering
\includegraphics[width=\linewidth]{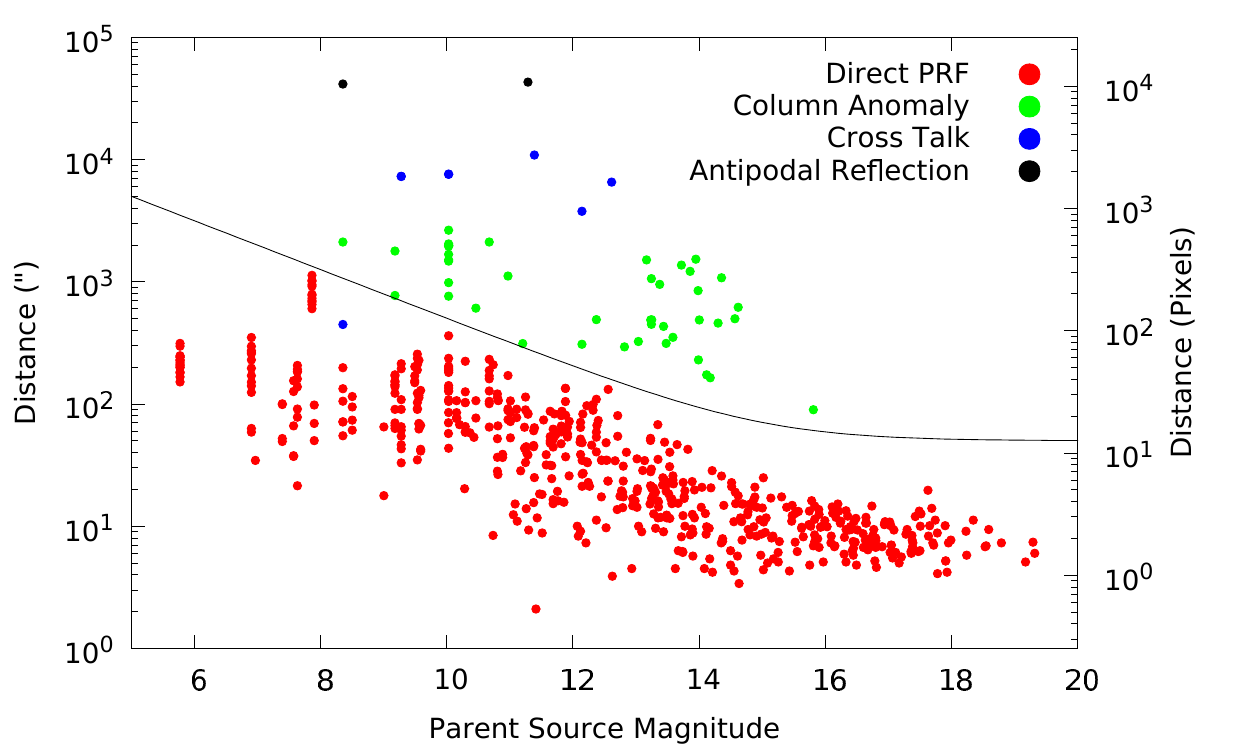}
\caption{The distance between each FP KOI child and its contaminating parent source. The mechanism of contamination for each FP KOI is shown via colored points. The solid line indicates the maximum distance at a given magnitude for the mechanism to be considered direct PRF contamination, as described by Equation~\ref{disteq}.}
\label{magdist}
\end{figure}

Criterion 1 ensured that no star was ever matched to itself. Criterion 2a is a semi-empirically determined formula derived to account for direct PRF contamination and reflection off the field flattener lens, assuming the average wings of a \emph{Kepler} PSF can be approximated by a Lorentzian distribution (see \S\ref{directprfsec}). The formula allows for any two stars to match within a generous 50$\arcsec$ (12.5 pixel) range, but allows for bright stars to match to larger distances, e.g., a 10$^{\rm th}$ mag star could match up to 500$\arcsec$ (125 pixels) away, and a 5th mag star could match up to 5000$\arcsec$ (1250 pixels) away. The factors of $50$ and $10^{6}$ were empirically adjusted to correspond to the natural boundary between the direct PRF and column anomaly matches, and we ensured that moderately increasing these limits did not yield additional, significant PRF matches. Figure~\ref{magdist} shows the distance between each FP KOI and its parent as a function of magnitude, along with the PRF contamination matching limit described by Equation~\ref{disteq} in criterion 2a. Criterion 2b accounts for antipodal reflection off the Schmidt Corrector (see \S\ref{reflectionsec}). Criterion 2c accounts for saturated/bleeding stars, CCD crosstalk (see \S\ref{crosstalksec}), and the column anomaly (see \S\ref{columncontamsec}). With respect to criterion 3, we match both A to B and B to A to ensure that any possible period ratio is found.

\begin{figure}
\centering
\includegraphics[width=\linewidth]{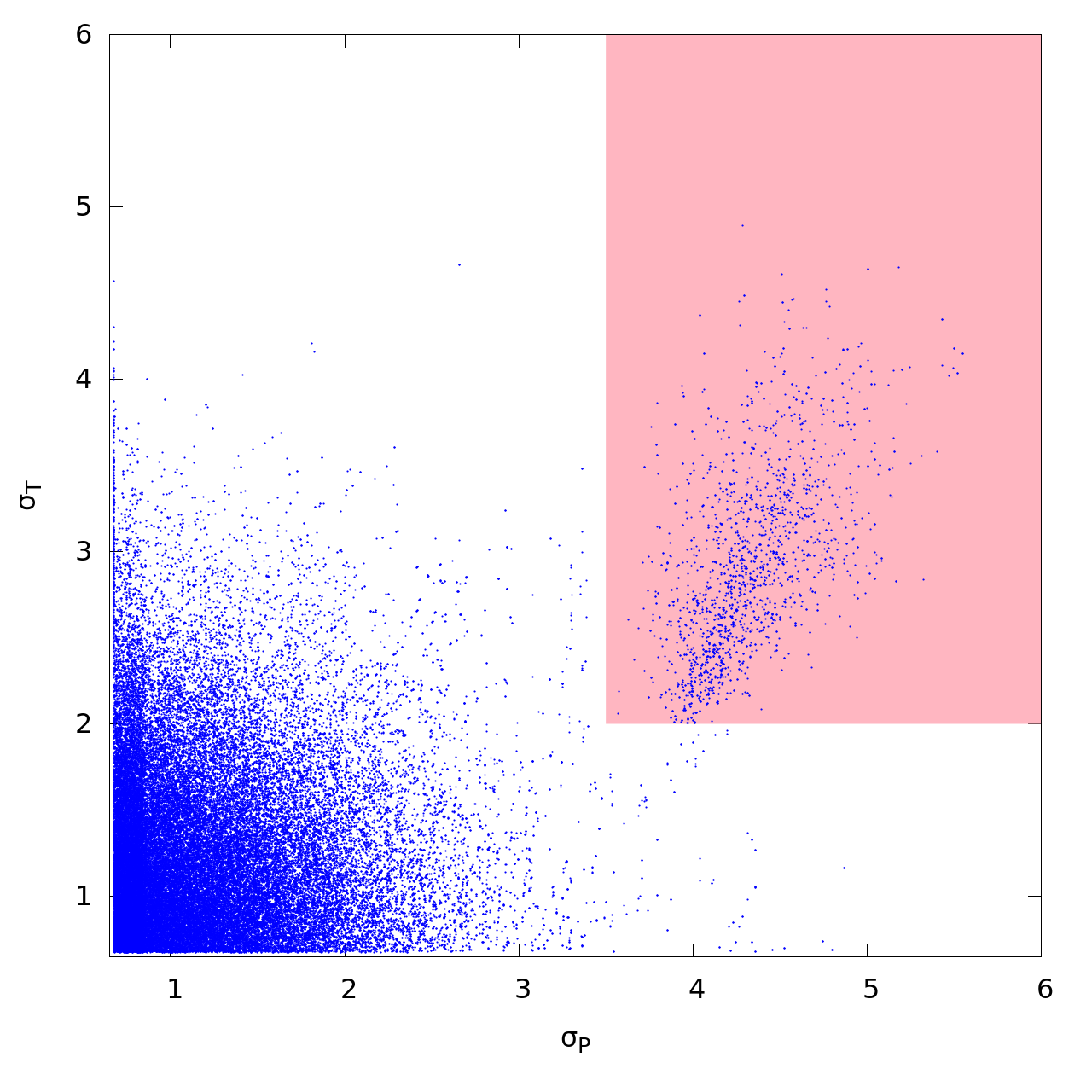}
\caption{A plot of all values for $\sigma_{P}$ and $\sigma_{T}$ when allowing period ratios of 1:1, 1:2, or 2:1 and employing physical constraints. The distribution of values most concentrated at ($\sigma_{P}$, $\sigma_{T}$) = (0.674, 0.674) correspond to random matches. The distribution centered at ($\sigma_{P}$, $\sigma_{T}$) = (4.5, 3.0) corresponds to significant, real, physically caused matches. The highlighted area that has its lower-left vertex at ($\sigma_{P}$, $\sigma_{T}$) = (3.5, 2.0) represents the parameter space in which a given match is considered significant.}
\label{sigfig}
\end{figure}

For criterion 3, the minimum $\sigma$ values were empirically determined to be $\sigma_{P,min}$ = 3.5 and $\sigma_{T,min}$ = 2.0, which allowed for a reasonable trade-off between capturing as many real matches as possible while excluding matches from random coincidence. Figure~\ref{sigfig} is a plot of $\sigma_{T}$ versus $\sigma_{P}$ for all possible matches at 1:1, 1:2, or 2:1 period ratios, while employing the physical constraints of criteria 1 and 2 above. As can be seen, a large, radial distribution of values, corresponding to completely random matches, is centered at ($\sigma_{P}$, $\sigma_{T}$) = (0.674, 0.674). This point represents the worst match possible with $\Delta P^{\prime}$ = $\Delta T^{\prime}$ = 0.5, and thus, for clarity, 0.674 = $\sqrt{2}\cdot \mathrm{erfcinv}(0.5)$. There is another distribution centered at ($\sigma_{P}$, $\sigma_{T}$) = (4.5, 3.0) that corresponds to the significant, real, physically caused matches. The pink highlighted area indicates the parameter space that satisfies criterion 3. The use of transit/eclipse duration was considered and investigated, but found to vary too much among the KOIs to be of consistent, practical use. Also, the sources employed for the GEB catalog did not all have eclipse duration measurements.

\subsection{Parent Source Determination}
\label{sourcedetsubsec}

While employing the aforementioned criteria, we performed a match of all KOIs to themselves, to all KEBs, and to all GEBs. Often for a given KOI, more than one significant match was found, as a parent often contaminates multiple stars. We were thus left with the task of identifying which object was the most likely physical source of the contamination, i.e., the parent. In general, given two matching objects, the one with the deeper transit/eclipse will be the most likely parent, as the known causes of contamination should always dilute the signal strength. Thus, for any given KOI, out of all the objects that matched and met the previously established criteria, the object with the deepest transit/eclipse was chosen as the most likely parent, and the KOI was deemed a FP. If there was no matching object that had a deeper transit/eclipse than a given KOI, then that KOI was not deemed a FP, and is likely a parent not a child.

Additionally, we also desired to identify the most likely physical mechanism of contamination, choosing among the four known mechanisms described in \S\ref{contammethsec}. If $d < d_{\rm max}$, as defined in Equation~\ref{disteq}, then the assumed mechanism was direct PRF contamination. If $d > d_{\rm max}$, and the matching stars were on different CCD modules, then the cause was determined to be due to antipodal reflection. If $d > d_{\rm max}$, the two stars were on the same CCD module, had different outputs, and had the same row and column number within ten pixels each, the mechanism was determined to be CCD crosstalk. If $d > d_{\rm max}$, they were on the same CCD module, and had the same column number within ten pixels, but different row numbers by more than ten pixels, then the mechanism was designated as column anomaly. Conversely, for completeness, if they had the same row number within ten pixels, but different column numbers by more than ten pixels, then the mechanism was designated as row anomaly. We note that no row anomalies were actually found, which lends confidence that the column anomaly is a real effect and that our matching criteria are yielding only statistically significant matches.

As there is a small region of known overlap between random matches and statistically significant matches (see Figure~\ref{sigfig}), we manually inspected the light curves of the matches most likely to be due to random chance. If the mechanism was anything other than direct PRF contamination, or if the period ratio was anything other than 1:1, 1:2, or 2:1, we compared the light curves to confirm they qualitatively shared the same morphology. For example, the transit/eclipse durations should roughly match, or if the parent has a secondary eclipse nearly as large as the primary, it should also be visible in the child. Although the vast majority of examined matches indeed had matching light curves, six did not: KOI~476.01 \& KOI~3673.01, KOI~982.01 \& KEB~002580872-sec, KOI~1943.01 \& KEB~006431670, KOI~2213.01 \& KEB~008572936-pri,  KOI~2220.04 \& KEB~006283224-pri, and KOI~3061.02 \& KOI~851.01. These cases all had values of ($\sigma_{P}$, $\sigma_{T}$) just above the cutoff value of (3.5, 2.0), and extreme period ratios, as expected for contamination from the vast population of random matches. These matches 
were thus eliminated from all tables and the KOIs were not designated as FPs.

\subsection{Modeling Direct PRF Contamination and Identifying Bastards}
\label{bastardkois}

One additional possibility we examined was that we had observed direct PRF matches between siblings, i.e., two children from the same parent, but did not observe the parent. For example, if two KOIs match, have nearly equal magnitudes and transit depths, but are separated by several arcseconds or more, neither can plausibly be the parent of the other. In these cases the parent must not be observed, either due to not being downloaded or not being located on a CCD, and we are thus observing bastards, i.e., children that are both contaminated by the same unobserved parent. In order to detect these bastards, we developed a simple model to describe the relation between the relative transit depths, magnitudes, and distances between two objects. If this model is significantly violated, it indicates that such a match has occurred, and thus the parent is unobserved and the two objects are bastards.

We start with the simple relations that the depths of the two objects, $D_{1}$ and $D_{2}$, are

\begin{equation}
\label{deptheq}
D_{1} = \frac{\Delta F_{1}}{F_{1}}
\qquad
D_{2} = \frac{\Delta F_{2}}{F_{2}}
\end{equation}

\noindent where $F_{1}$ and $F_{2}$ are the observed fluxes, and $\Delta F_{1}$ and $\Delta F_{2}$ are the observed changes in flux during transit, of the contaminated child and the proposed contaminating parent, respectively. (For clarity, object 1 is the child, and object 2 is the parent.) If we assume that $f_{\rm prf}(d)$ is a function that relates the fraction of object 2's flux that falls in the aperture of object 1 as a function of distance, then

\begin{equation}
\Delta F_{1} = \Delta F_{2} \cdot f_{\rm prf}(d)
\end{equation}

\noindent and

\begin{equation}
F_{1} = F_{1}' + F_{2} \cdot f_{\rm prf}(d)
\end{equation}

\noindent where $F_{1}'$ is the uncontaminated flux of object 1 such that

\begin{equation}
\label{fluxeq}
F_{1}' = F_{2} \cdot 10^{-0.4*\Delta m}
\end{equation}

\noindent where $\Delta m$ = $m_{\rm kep,1}$ - $m_{\rm kep,2}$. Combining Equations~\ref{deptheq}-\ref{fluxeq} yields the desired model relating the relative transit depths of a match to their relative magnitudes and distance,

\begin{equation}
\label{contameq}
\frac{D_{2}}{D_{1}} = 10^{-0.4\Delta m}\cdot f_{\rm prf}(d)^{-1} + 1
\end{equation}

With respect to the first term of Equation~\ref{contameq}, if the child (object 1) is much brighter than the parent (object 2), then the depth ratio will necessarily have to be large since any signal from the parent will be small compared to the flux of the child. Conversely, if the child is faint compared than the parent then the depth ratio could be close to unity since the child contributes very little extra flux to the total signal. With respect to the second term of Equation~\ref{contameq}, if the child and parent are very far apart, then the depth ratio will necessarily have to be large since very little flux from the parent will be in the aperture of the child. Conversely, if the child and parent are very close together, then the depth ratio can be close to unity as a very large fraction of the parent's flux will fall in the child's aperture.

We choose to represent the average \emph{Kepler} PRF as a combination of Gaussian and Lorentzian distributions, such that

\begin{equation}
\label{prfeq}
f_{\rm prf}(d) = \frac{1}{2}\exp\left(-\left(\frac{d^{2}}{\alpha^{2}}\right)\right) + \frac{1}{2}\left(\frac{\gamma^{2}}{d^{2}+\gamma^{2}}\right)
\end{equation}

\noindent where $\gamma$ is the half width at half maximum of the Lorentzian, and $\alpha$ is the standard deviation of the Gaussian. We choose the Gaussian and Lorentzian components because they adequately represent the PRF near each star, and the gradual wings of the PRF combined with reflection off the field flattener lens, respectively. We note that the \emph{Kepler} PRF is a very complicated distribution and heavily dependent on field location, and that the exact distribution of field flattener ghosts are not well understood, so this is a very simple approximation. We do not normalize the Gaussian nor Lorentzian functions because the parent should completely fill the child's photometric aperture when they are at the same location, i.e., $f_{\rm prf}$ = 1 at $d$ = 0.

We performed a robust fit of the model described by Equations~\ref{contameq}~and~\ref{prfeq} to all of the parent-child pairs identified as due to direct PRF contamination, utilizing their depth ratios, magnitude differences, and distances. After iteratively rejecting outliers greater than 3.2 times the standard deviation (where less than one outlier is expected for 700 data points) the fit converged with values of $\alpha$ = 6.73\arcsec and $\gamma$ = 0.406\arcsec. Outliers greater than 3.2 times the standard deviation of the final iteration, with these resulting fit parameters, were labeled as bastards. If the match involved two KOIs then both were designated as bastards and FP KOIs. In total, \nfpkois FP KOIs were identified, with \nbastards of them designated as bastards.

\subsection{Ephemeris Matching Results}

\begin{figure*}
\centering
\includegraphics[width=\linewidth]{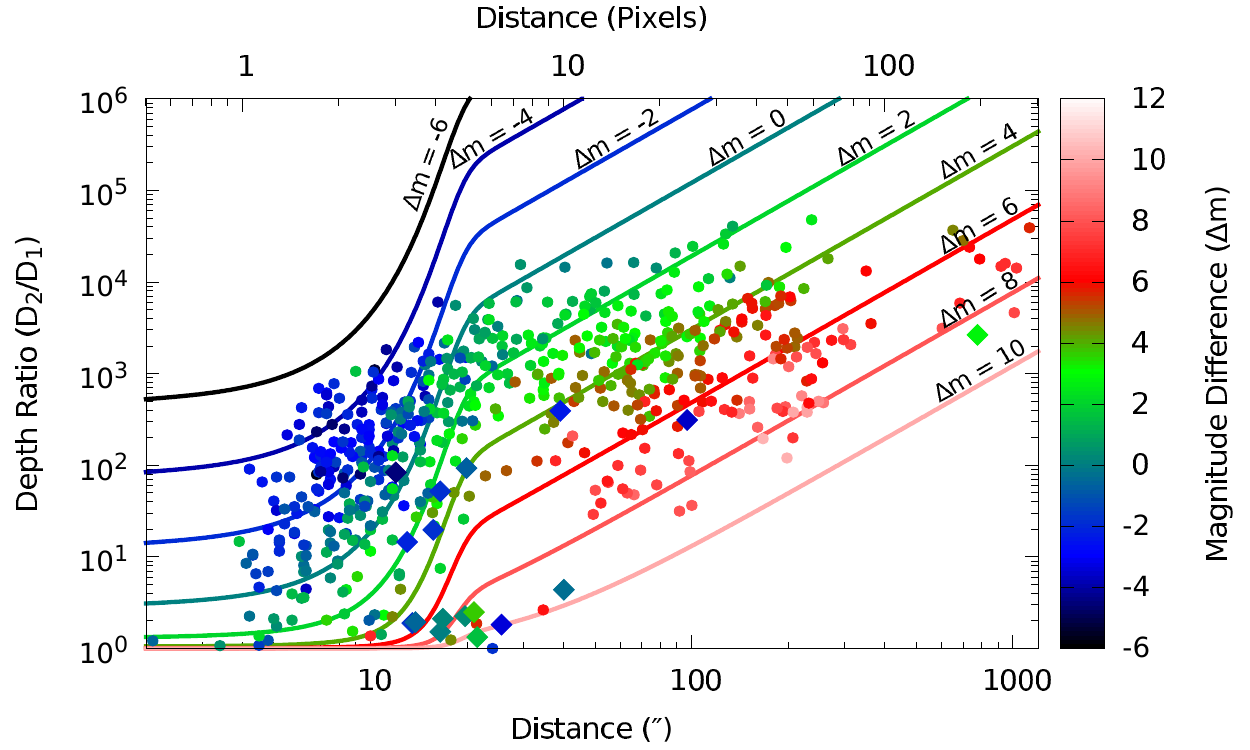}
\caption{A plot of the depth ratio between each FP KOI child and its parent, as a function of their distance and magnitude difference, for the Direct PRF contamination mechanism. The solid lines represent the best-fit model the data as represented in Equation~\ref{contameq}.}
\label{prfmodelfig}
\end{figure*}

In Figure~\ref{prfmodelfig} we plot the depth ratio, magnitude difference, and distance between each FP KOI and its most likely parent. Each pair is represented by a colored dot, where we have chosen color to represent magnitude difference. Solid colored lines outline the model represented by Equations~\ref{contameq}~and~\ref{prfeq}, at various intervals of magnitude difference, with the aforementioned best-fit parameters. Bastards are identified by larger diamond points.

In Table~\ref{matchtab} we list the \nfpkois KOIs we found to be FPs via ephemeris matching, grouped according to the mechanism of contamination. For each FP KOI we also list its KIC ID, the name and KIC ID of the most likely parent, the distance between the objects in arcseconds, the offset in row and column between the objects in pixels, the magnitude of the parent, the difference in magnitude between the KOI and the most likely parent, the depth ratio of the KOI and most likely parent, and a flag to designate unique situations.

\begin{deluxetable*}{rcrccrrrrrrc}
\tablecolumns{12}
\tablewidth{0in}
\tablecaption{The \nfpkois FP KOIs, Parents, and Properties}
\tablehead{\multicolumn{2}{c}{FP KOI (1)} & \multicolumn{2}{c}{Parent (2)} & Period & & & & & & \colhead{Depth} & Flag\\\colhead{KOI} & KIC & \colhead{Name} & KIC & Ratio & \colhead{Dist.} & \colhead{$\Delta$Row} & \colhead{$\Delta$Col} & \colhead{m$_{2}$} & \colhead{$\Delta$Mag} & \colhead{Ratio} & \\ & & & & (P$_{1}$:P$_{2}$) & \colhead{($\arcsec$)} & \colhead{(R$_{1}$-R$_{2}$)} & \colhead{(C$_{1}$-C$_{2}$)} & &  \colhead{(m$_{1}$-m$_{2})$} & \colhead{(D2/D1)} & }
\startdata
\\
\cutinhead{Antipodal Reflection}\\
3900.01 & 011911580 & 003644542-sec & 003644542 & 3:1 & 41461.2 & -18 & 44 & 8.35 & 5.55 & 2.41E+02 & 0\\
4646.01 & 012012439 & 003439031-sec & 003439031 & 1:2 & 42997.7 & 5 & 64 & 11.29 & 4.04 & 3.67E+03 & 0\\
\cutinhead{CCD Crosstalk}\\
559.01 & 006422367 & 005343976-pri & 005343976 & 1:1 & 6526.0 & 0 & 1 & 12.61 & 2.18 & 9.23E+02 & 0\\
1192.02 & 003644071 & 3511.01 & 003644542 & 3:1 & 447.0 & -6 & -1 & 8.35 & 5.86 & 6.14E+01 & 0\\
2908.01 & 006612284 & 006206751-pri & 006206751 & 1:1 & 3776.4 & 0 & -2 & 12.14 & 3.75 & 9.37E+02 & 0\\
\nodata & \nodata & \nodata & \nodata & \nodata & \nodata & \nodata & \nodata & \nodata & \nodata & \nodata & \nodata\\
\cutinhead{Column Anomaly}\\
1924.01 & 005108214 & 3688.01 & 005020034 & 1:1 & 89.7 & -22 & 1 & 15.81 & -7.97 & 2.47E+03 & 0\\
2233.01 & 008963721 & 009101279-pri & 009101279 & 1:1 & 1527.8 & 384 & 0 & 13.95 & 0.73 & 4.73E+03 & 0\\
2600.01 & 009777251 & BR-Cyg-pri & 009899416 & 1:1 & 1984.9 & 499 & 3 & 10.03 & 5.03 & 5.31E+03 & 0\\
\nodata & \nodata & \nodata & \nodata & \nodata & \nodata & \nodata & \nodata & \nodata & \nodata & \nodata & \nodata\\
\cutinhead{Direct PRF}\\
6.01 & 003248033 & 1759.01 & 003248019 & 1:1 & 14.3 & -4 & 0 & 15.39 & -3.23 & 1.69E+02 & 0\\
8.01 & 005903312 & 3692.01 & 005903301 & 1:1 & 8.0 & -2 & -1 & 15.15 & -2.70 & 2.67E+01 & 0\\
11.01 & 011913073 & 011913071-pri & 011913071 & 1:1 & 34.9 & 7 & -5 & 9.53 & 3.96 & 2.49E+02 & 0\\
\nodata & \nodata & \nodata & \nodata & \nodata & \nodata & \nodata & \nodata & \nodata & \nodata & \nodata & \nodata

\enddata
\tablecomments{Table~\ref{matchtab} is published in its entirety in the electronic edition of the Astronomical Journal. A portion is shown here for guidance regarding its form and content. A flag of 1 means the listed parent is likely not the true physical parent, but another contaminated object due to the same unobserved physical parent. A flag of 2 indicates that the computed depth of the KOI is anomalously large due to a bad crowding correction value or extreme quarter-to-quarter depth variations. A flag of 3 indicates that the given column anomaly occurs on another output of the same module, instead of the same output.}
\label{matchtab}
\end{deluxetable*}

Bastard FP KOIs are given a flag of ``1" in Table~\ref{matchtab}. The process of searching for bastards also revealed a few KOIs where the measured depth of the transit was significantly overestimated. We traced these cases back to bad crowding values or transit depths with large variations, and gave these matches a flag of ``2" in Table~\ref{matchtab}. Finally, there were a few cases where a match was identified as due to the column anomaly, except that the parent and child were located on different CCDs within the same module. While we are convinced these cases are significant and a real mechanism exists, we differentiate them from the more typical cases of column anomaly by giving them a flag of ``3" in Table~\ref{matchtab}.

\begin{figure*}
\centering
\includegraphics[width=\linewidth]{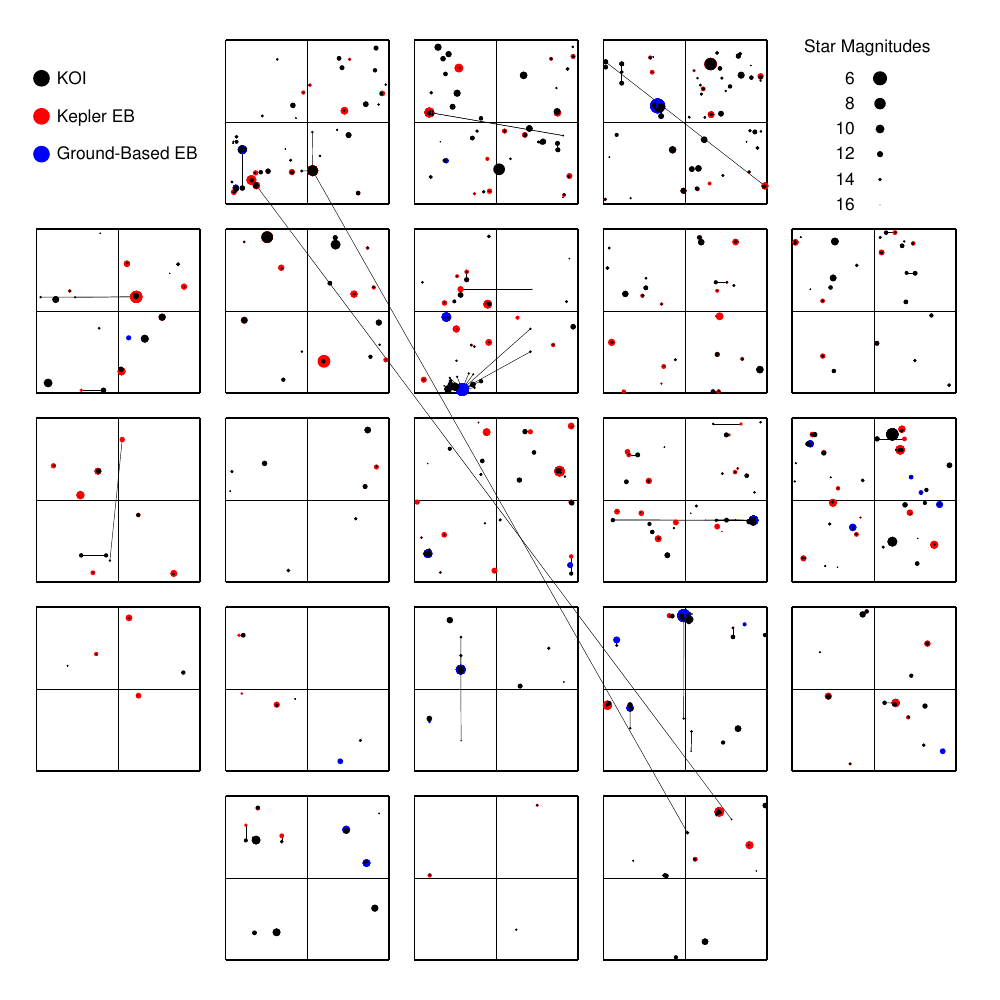}
\caption{Plot of the location of each FP KOI and its corresponding most likely physical parent source, connected by a solid black line. KOIs are represented by black points, KEBs are represented by red points, and GEBs are represented by blue points. The \emph{Kepler} magnitude of each star is shown via a scaled point size as indicated. Note that most parent-child pairs are so close together that the line connecting them is not easily visible on the scale of the plot. Also note that the sizes of the CCDs and their relative layout are not exact.}
\label{ccdplot}
\end{figure*}

In Figure~\ref{ccdplot} we plot the location of each FP KOI and its most likely parent, connected by a solid line. KOIs are represented by black points, KEBs are represented by red points, and GEBs are represented by blue points. The \emph{Kepler} magnitude of each star is shown via a scaled point size. Note that most parent-child pairs are so close together that the line connecting them is not easily visible on the scale of the plot.

\section{Discussion}
\label{discusssec}

Of the \nkois KOIs currently known at the time of this writing, we have deemed \nfpkois of them to be false positives, or 12\% of all known KOIs. In this section we discuss the properties of the FP KOIs and their parents, the true occurrence rate of FP KOIs due to contamination, and compare our results to other methods of detecting FP KOIs.

\subsection{Characteristics of FP KOIs}
\label{fpkoiprops}

\begin{figure*}
\centering
\begin{tabular}{cc}
\includegraphics[width=0.49\linewidth]{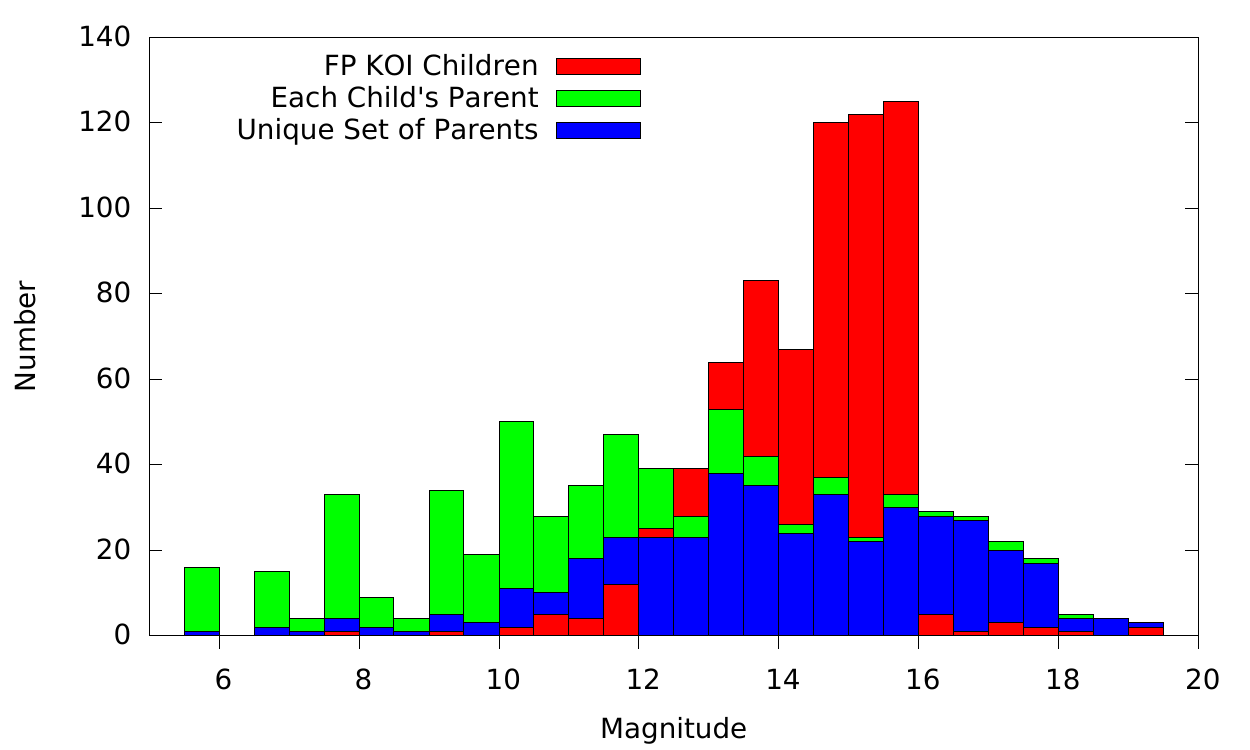} &
\includegraphics[width=0.49\linewidth]{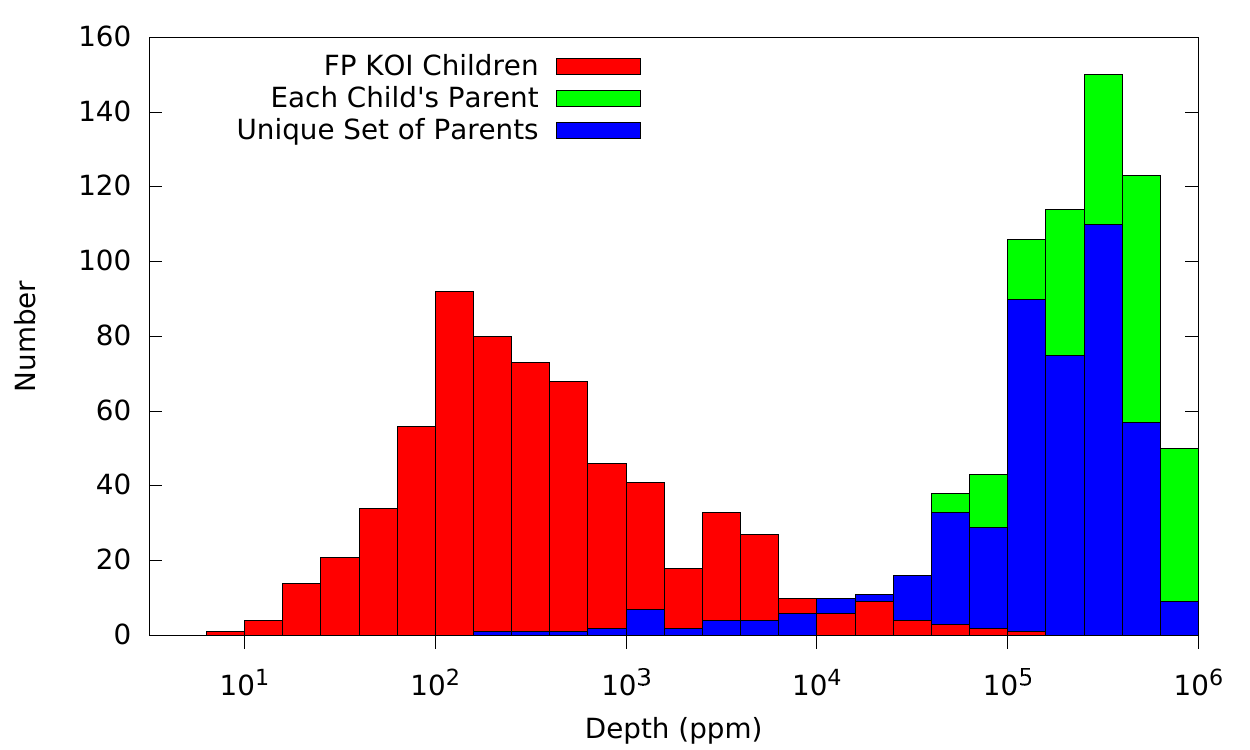}
\end{tabular}
\caption{Histogram plots of the magnitude (left) and depth (right) for each FP KOI, the parent of each FP KOI, and then each of the \nuniqsource unique parents. FP KOI children are shown in red, each child's parent is shown in green, and the unique set of parents is shown in blue. In order to show the entirety of these overlapping distributions, groups with lower values in a given histogram bin are shown in front of groups with higher values.}
\label{statfig1}
\end{figure*}

In Figure~\ref{statfig1} we plot histograms of the magnitude and depth for each FP KOI, the parent of each FP KOI, and then each of the \nuniqsource unique parents. (For clarity, in the second case we count each parent for every FP KOI it spawns, but in the last case we only count each parent once.) FP KOI children are shown in red, each child's parent is shown in green, and the unique set of parents is shown in blue. In order to show the entirety of these overlapping distributions we ensure that groups with lower values in a given histogram bin are shown in front of groups with higher values. From this figure, it can be seen that a few bright parents with deep eclipses are responsible for a significant number of FP KOIs. The FP KOI population peaks just brighter than 16$^{\rm th}$ magnitude, although there is an artificial cut off at greater values because targets fainter than 16$^{\rm th}$ magnitude were not included on the initial primary mission target list. FP KOI depths reach as small as 10 ppm, and thus encompass the depth of an expected Earth-analogue at 84 ppm.

\begin{figure*}
\centering
\begin{tabular}{cc}
\includegraphics[width=0.49\linewidth]{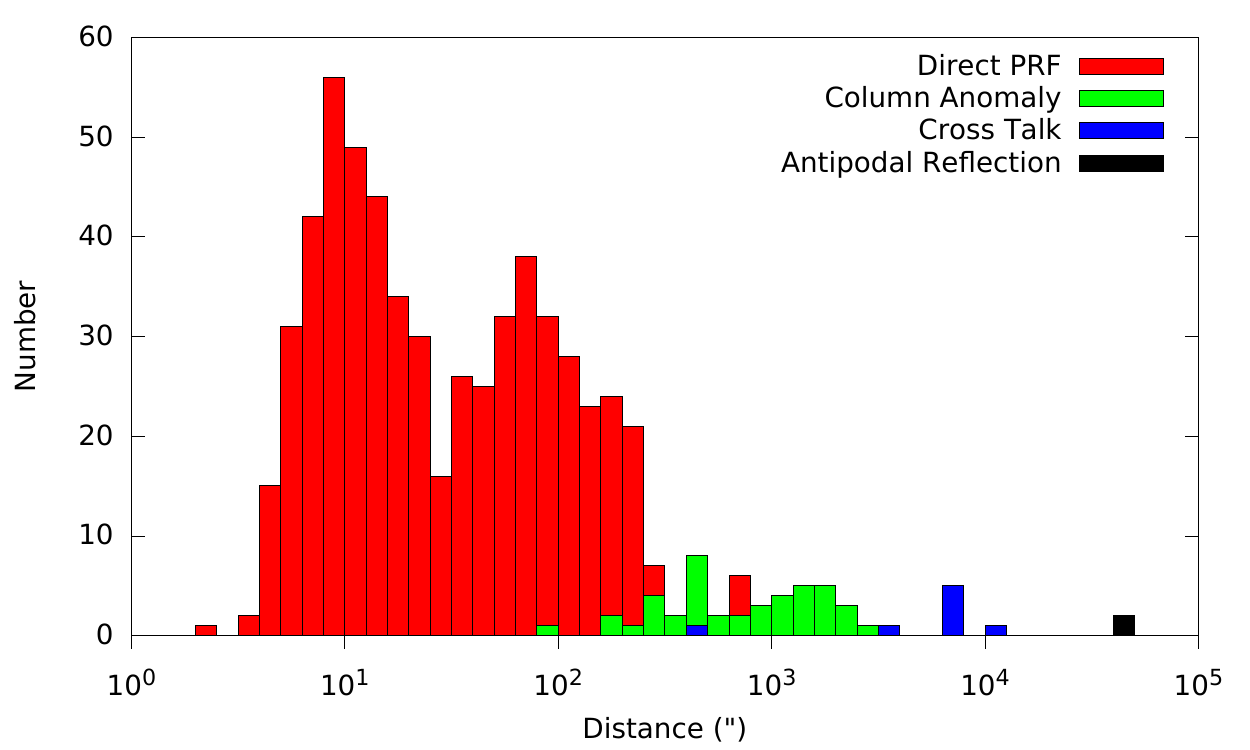} &
\includegraphics[width=0.49\linewidth]{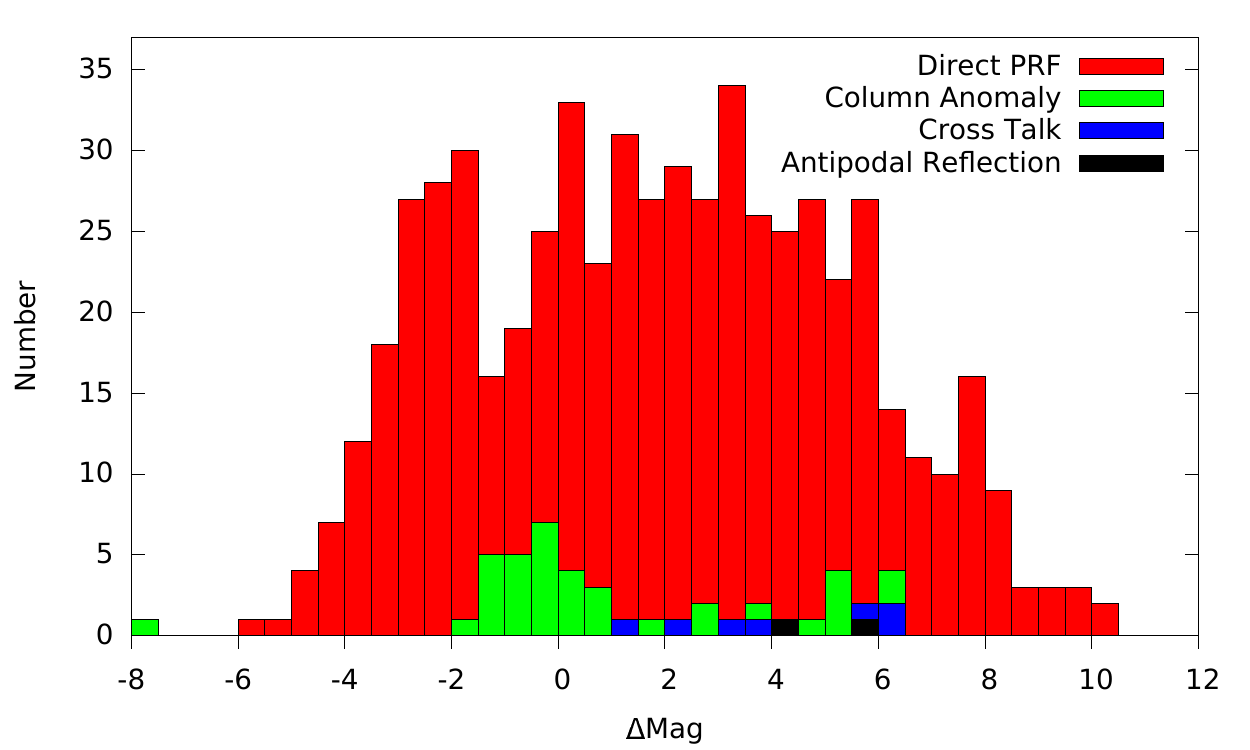} \\
\includegraphics[width=0.49\linewidth]{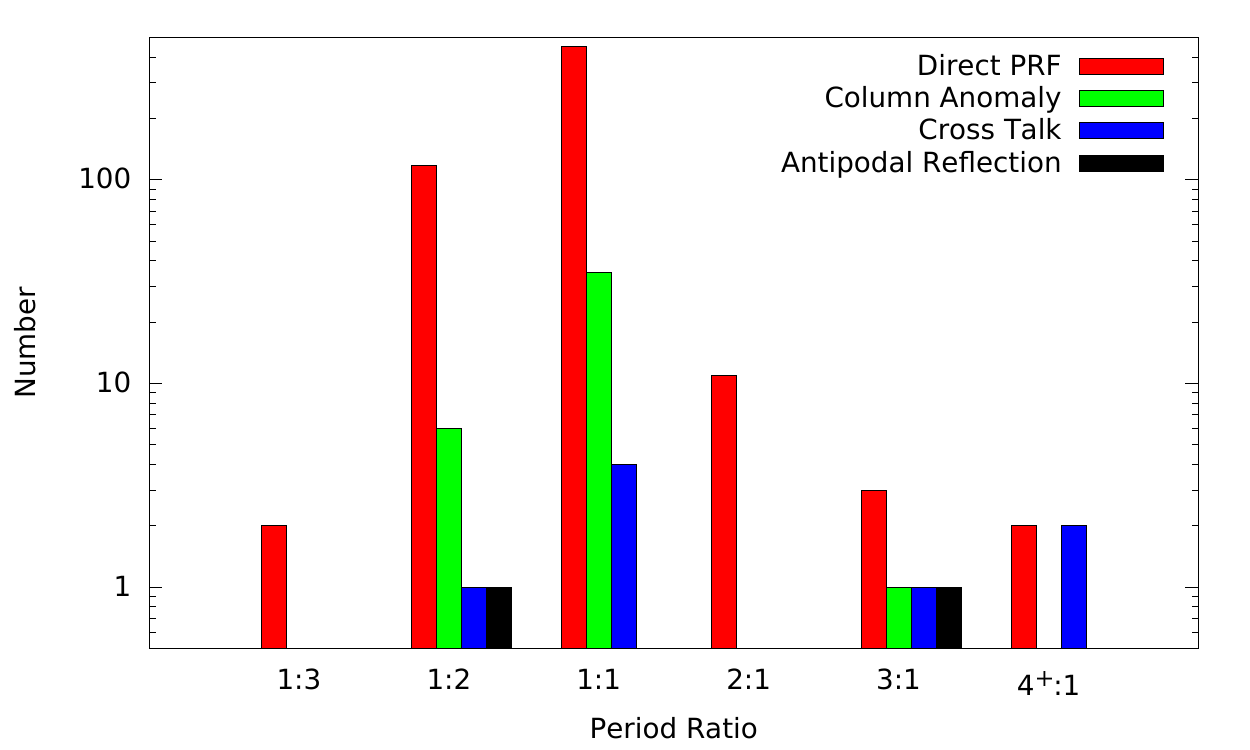} &
\includegraphics[width=0.49\linewidth]{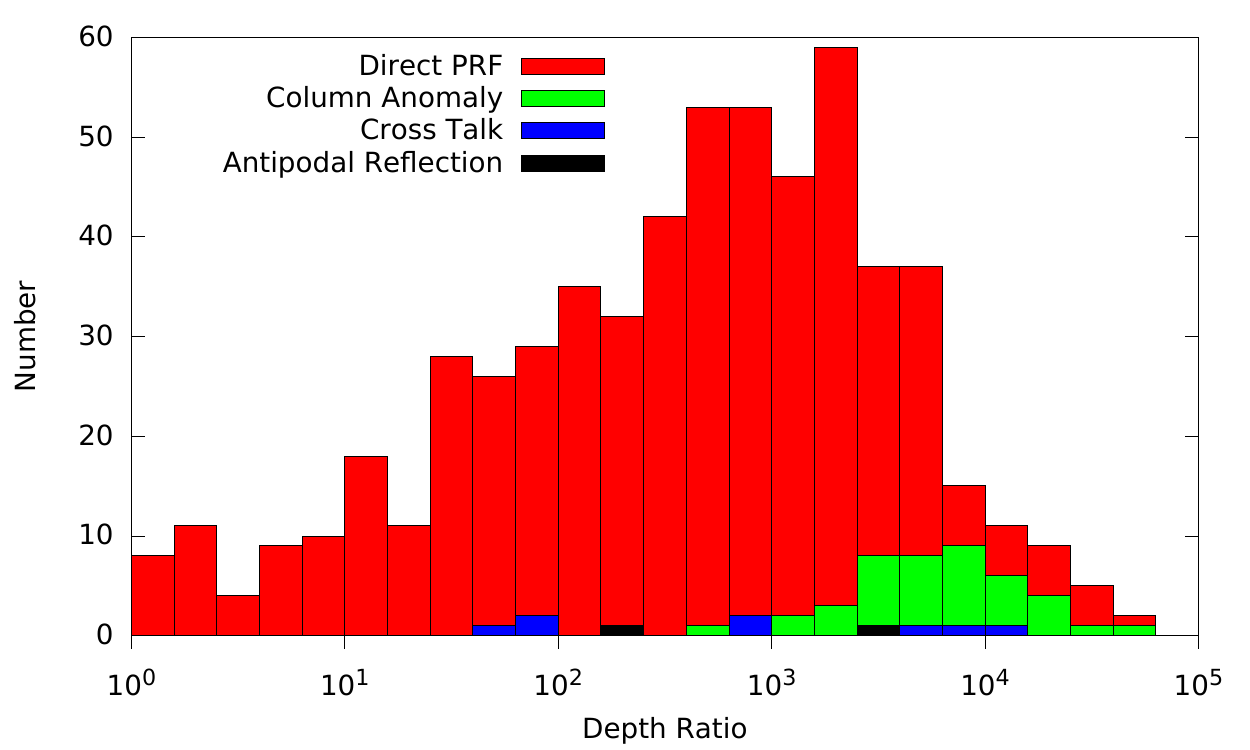} \\
\end{tabular}
\caption{Histogram plots of the distance (top-left), relative magnitude (top-right), period ratio (bottom-left), and depth ratio (bottom-right) for each FP KOI and its corresponding most likely parent. The different mechanisms of Direct PRF, Column Anomaly, CCD Crosstalk, and Antipodal Reflection are represented by red, green, blue, and black colors respectively. Only matches with a flag of 0 in Table~\ref{matchtab} are included in these plots.}
\label{statfig2}
\end{figure*}

In Figure~\ref{statfig2} we plot histograms of various parameters between FP KOIs and their parents. This includes the distance between each parent and child, their magnitude difference, their depth ratio, and their period ratio. The four mechanisms of contamination are indicated by different colors. Only matches with a flag of 0 in Table~\ref{matchtab} are included in these plots, as we did not want the statistics to be biased by bastards, wildly incorrect depths, or the few outlying cases where column anomaly is occurring between two different CCDs on the same module.

Examining the distribution of the distances between parents and children, most Direct PRF contamination occurs when the separation is less than $\sim$300$\arcsec$. There is a bimodal distribution with peaks at $\sim$10$\arcsec$ and $\sim$100$\arcsec$. The first peak likely corresponds to the majority of stars that are faint and can only contaminate other stars that are within $\sim$10-20$\arcsec$. The second peak likely corresponds to the handful of very bright variable stars in the field, which can each spawn tens of FP KOIs out to hundreds of arcseconds. Ghost images from reflection off the field flattening lenses likely compose a significant fraction of this second peak. As expected, the other three mechanisms contaminate to much further distances, but are drastically smaller in number.

Examining the distribution of differential magnitude between matches, contamination appears to occur over a very wide range. The vast majority of matches occur for -5 $<$ $\Delta$m $<$ 10, i.e., the child can be up to 5 magnitudes brighter than its parent, or down to 10 magnitudes fainter than its parent. For 67\% of the FP KOIs the parent is brighter, but for 33\% of them the parent is fainter. For column anomalies it appears that the parent is either very close to the same magnitude of the FP KOI, or much brighter than it. Cross-talk spans a range of magnitudes, though the parent is always brighter, and for both cases of reflection the parent was $\sim$5 magnitudes brighter.

Examining the distribution of period ratios, 77\% of all FP KOIs have the same period (1:1 ratio) as their parent, as expected. Another 19\% have periods half that of their parent (1:2 ratio). These are due to EBs with nearly equal primary and secondary eclipse depths so that the resulting FP KOI is detected at half the binary period. The remaining 4\% of period ratios consist of unusual cases. For example, the two FP KOIs with periods one-third their parents (1:3 ratios) both had parents that were eccentric EBs with secondary eclipses located at a phase of 0.667. This allows for the binary to be folded at one-third of the true period and have the primary and secondary eclipses stack on top of each other, producing a FP KOI at one-third the binary period. Cases where the period of the FP KOI is twice their parent (2:1 ratio) and greater are typically due to varying levels of contamination quarter-to-quarter that cause the FP KOI signal to only be present in selected quarters. Direct PRF contamination can vary due to changing optimal apertures and PRF distributions. The column anomaly can vary as the stars fall on physically different pixels each quarter, and both parent and child may not fall on the same column due to imperfect CCD registration. Cross-talk varies heavily as it is strongly CCD dependent and the stars fall on different CCDs each season. Finally, antipodal reflection can vary due to small offsets in the position of the \emph{Kepler} boresight. When quarter-to-quarter variations exist, and are coupled with longer orbital periods so that only a few transits or less are visible each quarter, large period ratios can be produced. The most extreme period ratios found were the FP KOIs 3827.01 and 3827.02, which were 13 and 15 times the period of their common parent, KEB 003858884. This is a 25.9 day, 9.27 $m_{\rm kep}$ EB which contaminated via cross-talk only every fourth quarter, with a few eclipses not observed due to data gaps.

Examining the depth ratio between matches, the distribution peaks at $\sim$$10^{3}$, when plotted in log space, with a gradual decline toward smaller values, and a steep decline towards larger values. The drop off at depth ratios close to unity is due to the requirement that the parent must be right next to, and much brighter than, the child in order to induce a similar depth, which happens rarely. The drop off at large depth ratios is likely due to observational detection bias. If we assume the smallest signal we could detect is 10 ppm, then at a depth ratio of $10^{5}$ the parent would have to be an EB with a nearly total (100\%) primary eclipse. There are expectedly many cases of contamination where the resulting signal is below the detection threshold, and for long periods and/or faint stars the detection threshold is certainly at least an order of magnitude higher than 10 ppm. At some flux level, every star in the \emph{Kepler} field is contaminated. As more data is analyzed, e.g., the 17 quarters of data now collected by \emph{Kepler}, smaller transit depths and higher depth ratios will be probed, and thus more false positives due to contamination will certainly be found.

\subsection{FP KOI Prevalence and Occurrence Rate}
\label{ratesec}

In order to define a FP KOI status with ephemeris matching, both the parent and the child, or two siblings, must be observed. Of the \nfpkois FP KOIs we identified, we found that for \nbastards of them, or 4.5\%, we were not able to identify a physically plausible parent. There are certainly many more FP KOIs that exist in the field that we are not able to identify via ephemeris matching because we do not observe the parent nor another matching child.

Since only 34\% of all stars in the field are downloaded, we would expect, to a first order approximation, that we would only observe the parent of a FP KOI, and thus have it show up as an ephemeris match to another KOI or EB, 34\% of the time. This means that for every FP KOI we have found via ephemeris matching, there are approximately another two KOIs that are FPs due to contamination, i.e., we are only 34\% complete in identifying FPs via the ephemeris matching method. Compensating for this bias would raise the overall FP rate of KOIs due to contamination from 12\% to 35\%. 

This is a very simple approximation because the stars that \emph{Kepler} observes are not selected randomly. The \emph{Kepler} sample is more complete for bright stars, i.e., it observes most of the bright stars in the field, than it is for faint stars. However, it contains far greater numbers of fainter stars, as there are simply a far greater number of faint stars in the field\citep{Batalha2010}. As discussed in \S\ref{fpkoiprops}, the parent of an FP KOI child, based on the current observations, is twice as likely to be brighter than the child than fainter. Brighter stars are capable of producing many more FP KOIs than fainter stars, as evidenced by the fact that the \nfpkois FP KOIs identified in this paper are caused by only \nuniqsource unique parent sources. Thus, \emph{Kepler} is observing a greater fraction of the bright stars that are more prone to cause FP KOIs, but there are far greater numbers of faint stars that we do not observe. A careful statistical analysis is required to exactly determine the effect that these two competing biases has on the determination of the true false positive rate for \emph{Kepler} KOIs. While this is unfortunately beyond the scope of this current paper, it will hopefully be addressed in follow-up studies, utilizing even more \emph{Kepler} data.

It is of further interest to examine the contamination rate with respect to multi-planet systems. We examined \nkois KOIs, of which \nmultikois have more than one KOI assigned to the same star, i.e., are multi-planet candidates. Of the \nfpkois FP KOIs, \nfpmultis are part of multi-planet systems. This means that for the single-planet candidate KOI population, we observed a 16\% FP occurrence rate due to contamination, but only an observed 2.4\% FP rate for the multi-planet KOIs. This lends further credence to other studies that the false positive rate of planet candidates in multi-planet systems is much lower than for lone candidates, e.g., \citet{Lissauer2013} and \citet{Rowe2013}.

\subsection{Comparison to FP KOIs Detected Via Other Methods}
\label{tcertsec}

As discussed in \S\ref{introsec}, FP KOIs are often detected either via an evident secondary eclipse in the light curve, or an observed offset in the location of the transit signal in the pixel-level data \citep{Bryson2013}. These methods often work well when the KOI has a high SNR and/or when the parent is close by, but can become ineffective when applied to KOIs with low SNR and/or distant parents. As part of the \emph{Kepler} project, the threshold crossing event review team (TCERT) is responsible for evaluating every KOI using these flux and pixel-level techniques, and determining if the KOI is a FP. It is thus of interest to compare the results from TCERT to our results.

\begin{figure}
\centering
\includegraphics[width=\linewidth]{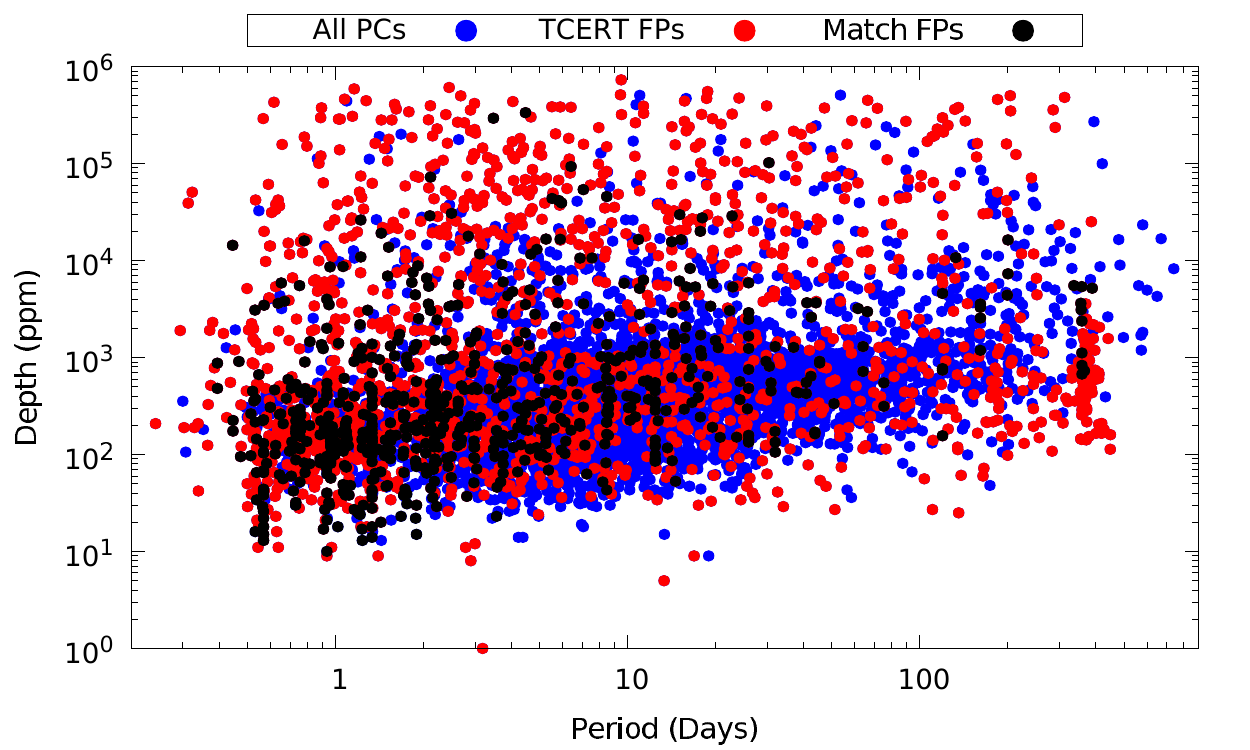}
\caption{A plot of transit depth vs period for KOIs that are planetary candidates (PCs - blue circles), false positives designed by TCERT (TCERT FPs - red circles), and false positives designed using the ephemeris matching technique in this paper (Match FPs - black circles).}
\label{pdplot}
\end{figure}

Of the \nfpkois FP KOIs identified in this paper, 352 were designated as KOIs based on analysis of 8 quarters or less of \emph{Kepler} data \citep{Borucki2011a,Borucki2011b,Batalha2013,Burke2013}. Of this older group, 327 have already been determined to be FPs by TCERT, leaving 25 new FP KOIs from this group that were identified as a result of the ephemeris matching technique in this paper. However, the other 333 FP KOIs were designated as KOIs based on 12 quarters of \emph{Kepler} data \citep[][in preparation]{Rowe2014}, and only 240 of these were identified by TCERT as FPs, leaving 93 as newly identified FPs. Altogether this means that 118 new FP KOIs have been identified as a result of this paper and the ephemeris matching technique. In Figure~\ref{pdplot} we plot transit depth vs period for KOIs that are planetary candidates, KOIs that are false positives designed by TCERT, and KOIs that are false positives designed using the ephemeris matching technique in this paper.

While TCERT was 92.9\% effective in detecting FPs for the Q8 and earlier KOIs, it was only 72.1\% effective with the newer Q12 KOIs, compared to the results in this paper. The first set of KOIs, based on 8 quarters of data or less, included transit-like signals with depths ranging from extremely deep to as shallow as the noise limit permitted. The new KOIs resulting from analyzing 12 quarters of data thus predominately included shallower transit-like signals that only became detectable with the addition of $\sim$50\% more data, as the deeper transit-like signals had already been found. (A small number of new long-period systems were also found at a range of transit depths with the extended temporal baseline.) Since the new KOIs from the Q12 analysis have shallower transit depths, they are both lower SNR and could be contaminated by a parent that is farther away. It is thus not surprising that most new FP KOIs were found to be among this latter group, as the TCERT diagnostics are less effective with low SNR transits and distant parents.

\subsection{Future Work}

At the time of this writing there are now 17 quarters of \emph{Kepler} data available, which will extend the detectable period range by 40\%, and the transit depth by 19\%, compared to using 12 quarters of data. As discussed in \S\ref{fpkoiprops} and \S\ref{tcertsec}, at lower transit depths the number of FP KOIs rises significantly and the methods currently employed by TCERT become less effective. While we have shown that ephemeris matching is a complementary method that is capable of detecting low SNR KOIs with small transit depths that other methods cannot, as discussed in \S\ref{ratesec}, ephemeris matching will only detect $\sim$34\% of FP KOIs. It is thus of paramount importance to be able to reliably detect FP KOIs. 

The addition of the GEB catalog helps mitigate the problem that only 34\% of stars are observed, by adding additional systems to find matches to that are not downloaded by \emph{Kepler}, but only to a certain extent. The compilation of ground-based catalogs has a peak distribution in magnitude of $\sim$13.5, and thus this sample of EBs is likely only complete to that magnitude, with very few GEBs known fainter than $\sim$16$^{\rm th}$ magnitude. Furthermore, EBs observed from the ground suffer from further biases that afflict ground-based surveys; they can only observe at night, when weather permits, and only for the part of the year the field is visible at night. Thus, most detached EBs discovered from the ground are at short orbital periods of $\sim$10 days or less. Since almost no long period GEBs are known, they are of limited use in identifying FPs that appear to be Earth-like planets in the habitable zone.

We thus suggest obtaining simple ground-based photometric observations of a KOI and its surrounding field out to several arcminutes when a transit is expected. Certainly one cannot expect to observe a signal of $\sim$84 ppm, as expected for an Earth-like candidate, from the ground. However, if the KOI is a FP due to contamination from an eclipsing binary or other variable star, as previously discussed, it is very likely that the parent has an eclipse depth of $>$ 1\%. A signal of this depth is easily detectable even with small telescopes equipped with CCD cameras, such as those possessed by universities and advanced amateur astronomers. An organized campaign by those with these modest resources could allow for the elimination of most contamination scenarios, and significantly bolster the confidence that a given KOI is a true transiting planet.

\acknowledgments
We thank the anonymous referee for his or her very helpful comments, which especially helped to improve the clarity of the paper. This research has made use of the NASA Exoplanet Archive, which is operated by the California Institute of Technology, under contract with the National Aeronautics and Space Administration under the Exoplanet Exploration Program. This research has made use of NASA's Astrophysics Data System. Some of the data presented in this paper were obtained from the Mikulski Archive for Space Telescopes (MAST). STScI is operated by the Association of Universities for Research in Astronomy, Inc., under NASA contract NAS5-26555. Support for MAST for non-HST data is provided by the NASA Office of Space Science via grant NNX09AF08G and by other grants and contracts. This paper includes data collected by the \emph{Kepler} mission. Funding for the \emph{Kepler} mission is provided by the NASA Science Mission directorate.

% Bibliography
\bibliography{AstroRefs.bib}

\end{document}